\documentclass[letter]{aa} 

\usepackage{graphicx}
\usepackage{txfonts}
\usepackage{soul}
\begin{document}

   
   \title{An Ultra diffuse Galaxy in the NGC~5846 group from the VEGAS survey}


   \author{Duncan A. Forbes
          \inst{1}, Jonah Gannon \inst{1}, Warrick J Couch \inst{1},  Enrichetta Iodice \inst{2},
          Marilena Spavone \inst{2}, Michele Cantiello \inst{3}, Nicola Napolitano \inst{4} \and Pietro Schipani \inst{2}
         }

   \institute{Centre for Astrophysics \& Supercomputing, Swinburne University, Hawthorn VIC 3122, Australia\\
              \email{dforbes@swin.edu.au}
         \and
             INAF-Astronomical Observatory of Capodimonte
Salita Moiariello 16, 80131, Naples, Italy
\and
INAF astronomical observatory of Abruzzo, via Maggini snc, 64100, Teramo, Italy
\and
School of Physics and Astronomy, Sun Yat-sen University Zhuhai Campus, 2 Daxue Road, Tangjia, Zhuhai, Guangdong 519082, P.R. China\\
             }

   \date{Received September 15, 1996; accepted March 16, 1997}

 
  \abstract
   {Many ultra diffuse galaxies (UDGs) have now been identified in clusters of galaxies. However, the number of nearby UDGs suitable for detailed follow-up remain rare.}
   {Our aim is to begin to identify UDGs in the environments of nearby bright early-type galaxies from the VEGAS survey. }
   {Here we use a deep g band image of the NGC 5846 group, taken as part of the VEGAS survey, to search for UDGs.}
   {We found one object with properties of a UDG if it associated with the NGC 5846 group, which seems likely. The galaxy, we name NGC~5846$\_$UDG1, has an absolute magnitude of M$_g$ = --14.2, corresponding to a stellar mass of $\sim$10$^8$ M$_{\odot}$. It also reveals a system of compact sources which are likely globular clusters. Based on the number of globular clusters detected we estimate a halo mass that is greater than 8$\times$10$^{10}$ M$_{\odot}$ for UDG1.}
   {}

   \keywords{Galaxies: dwarf  -- Galaxies: formation -- Galaxies: star clusters: general -- Galaxies: halos                   }

\authorrunning{Forbes et al.} 

   \maketitle
%

\section{Introduction}

Since the identification in 2015 of around 50 galaxies given the name `ultra diffuse galaxies' (UDGs) in the Coma cluster by van Dokkum et al. (2015), we now know of hundreds of UDG candidates in the direction of the Coma cluster (Yagi et al. 2016).
Spectroscopic followup of some of these candidates reveals a high fraction associated with the Coma cluster at a distance of $\sim$100 Mpc (Alabi et al. 2018).  With a UDG defined to have an effective radius R$_e$ $>$ 1.5 kpc and a central g band surface brightness $\mu_0$ $>$ 24 mag. sq. arcsec, they have been found in the full  range of galactic environments from relative isolation (Martinez-Delgado et al. 2016; Roman et al. 2019), to groups (Roman \& Trujillo 2016), to 
the richest clusters (Janssens et al. 2017). Further studies have revealed that the abundance of UDGs scales in a near-linear fashion with the halo mass of the group/cluster (Mancera Pina et al. 2018), so that a 10$^{13}$ M$_{\odot}$ group would be expected to host half a dozen to a dozen UDGs within its virial radius.

The number of nearby ($\le$ 30 Mpc) UDGs in the literature is still relatively small. Within the Local Group, the WLM
dwarf has a size and surface brightness consistent with that of a UDG (e.g. see figure 6 of Greco et al. 2018). Additional nearby UDG candidates, including those associated with the Cen A group (distance = 3.5 Mpc) are listed by Rong et al. (2017). 
Recently, Muller et al. (2018) identified 5 UDG candidates with effective radii $\ge$ 1.5 kpc if associated with the Leo group (distance 10 Mpc). To date, there has been no published distance confirmation for these UDG candidates. 
Trujillo et al. (2017) identified one UDG (UGC~2162) in the M77 group, which is located at a distance of 12 Mpc.
Two UDGs (DF2 and DF4) have been associated with the NGC 1052 group (distance $\sim$20 Mpc) based on surface brightness fluctuations (see Blakeslee \& Cantiello 2018 for a recent analysis) and both have been inferred to have little, or no, dark matter (van Dokkum et al. 2019). However, the distance, and the claim for the paucity of dark matter, are the subject of much debate in the literature (e.g. Trujillo et al. 2018). Merritt et al. (2016) has identified 4 very diffuse UDGs ($\mu_0 = 25.6-27.7$ mag. sq. arcsec in the g band) associated with the giant elliptical NGC 5845 (distance 27 Mpc). Finally, we note the work of Venhola et al. (2017) who used OmegaCAM on VST to search for UDGs in the Fornax cluster (distance = 20 Mpc, virial mass = 7 $\times$ 10$^{13}$ M$_{\odot}$). They found 9 UDG candidates.

Here we report a UDG found in the deep imaging of the NGC 5846 group as part of the VEGAS survey. Mahdavi et al. (2005) noted that the NGC 5846 group is relatively isolated in redshift space and is dominated by two large ellipticals: NGC 5846 (V = 1712 km/s, 
D = 25 $\pm$ 4 Mpc) and NGC 5813 (V = 1956 km/s, D = 32 $\pm$ 3 Mpc). The UDG is much closer in projection (21 arcmin) to NGC 5846 than NGC 5813, however we can't rule out association with the NGC 5813 subgroup. Nevertheless, Mahdavi et al. concluded that NGC 5846 and NGC 5813 were at a "compatible distance". Galaxies in the NGC 5846 group have a range of 900 $<$ cz $<$ 2700 km/s with a peak around 1900 km/s. 
Given it is projected near the centre of the NGC 5846 group, we assume it belongs to the main group and have named it NGC~5846$\_$UDG1, or UDG1 for short.  The NGC 5846 group has a halo mass of $8 \times 10^{13}$ M$_{\odot}$ (Mahdavi et al. 2005). 
After briefly describing the VEGAS survey, we present details of UDG1. Throughout this paper we assume the same distance to the NGC 5846 group as Spavone et al. (2017), i.e. 24.89 Mpc from the SBF distance measurement of Tonry et al. (2001). 

\section{VEGAS Imaging}

The ongoing VEGAS survey (http://www.na.astro.it/vegas)  is using OmegaCAM on the VST to image 100 bright nearby early-type galaxies in a range of environments. The survey focuses on both compact stellar systems in galaxy halos and the diffuse light within galaxy halos. 
A data reduction pipeline, called VST-Tube (Grado et al. 2012; Capaccioli et al. 2015), has been developed to create image mosaics that optimise the detection of low surface brightness features. 
The survey aims to obtain a systematic depth of 27.3 mag. per arcsec$^2$ with a S/N $>$ 3 per arcsec$^2$ in the g band. For further details of the survey and examples of depths reached see Capaccioli et al. (2015) and Spavone et al. (2017).

Dithered images of the NGC 5846 group were taken as part of the VEGAS survey on 2015 June 4 under seeing conditions of $\sim$1 arcsec. Deep g band images were taken with a total exposure time of 11250 sec. Shallower i band images (exposure time of 1650 sec) were also obtained but these images are not used for UDG detection as they are too shallow to reveal any UDGs. The multiple g band images were combined to give a final mosaic image of 1$\times$1 sq. degree (with 0.21 arcsec pixels). The final image has a
limiting surface brightness of 28.6 mag. per arcsec$^2$ in the g band (for a S/N $\sim$3 per arcsec$^2$). 

\section{An Ultra Diffuse Galaxy in the NGC 5846 Group}

The search for UDGs in the NGC 5846 group was conducted visually. Initially, the full 16K $\times$ 16K mosaic image was displayed on a high resolution 4K $\times$ 4K device. The mosaic was then examined in subsections so that individual pixels were resolved. The image was carefully examined visually for low surface brightness, large objects of diameter of tens of pixels (corresponding to several kpc at the distance of the NGC 5846 group). Although we identified several small objects (such as dE,N galaxies) we found only one low surface brightness, large object. An image of this object is shown in  Fig. 1.

This object appears to resemble an Ultra Diffuse Galaxy, which we call UDG1.
Also apparent in the image are a number of bright, compact objects which are concentrated around the UDG. These are likely to be globular clusters, although one of the brightest may be a nuclear star cluster (the galaxy centre is poorly defined). 

In order to measure the properties of UDG1 (and hence confirm its status as a UDG) we used the {\it ellipse} task within IRAF. This task fits isophotes of constant surface brightness at ever increasing radii. We fixed the centre to the bright star cluster near the galaxy centre. Other star clusters were excluded from the galaxy model using an iterative sigma clipping technique.  Our resulting model did not allow measurement of the galaxy ellipticity or position angle due to the low surface brightness. However, the galaxy appears fairly circular with no obvious signs of tidal interaction. We were able to derive a curve-of-growth style total magnitude.
Fitting the curve-of-growth beyond 2 arcsec, we measure a total magnitude of m$_g$ = 18.0 $\pm$ 0.2 or M$_g$ = --14. Correcting for Galactic extinction of A$_g$ = 0.16, this becomes M$_g$ = --14.2. 
From the VEGAS i band image, we estimate its global g--i colour to be $\sim$1, i.e. quite red, suggestive of an old stellar population with M/L$_g$ $\sim$ 3 (Into \& Portinari 2013). 
Its total luminosity corresponds to a stellar mass of $\sim$ 10$^8$ M$_{\odot}$. 
The curve-of-growth also allows us to estimate the radius that contains half of the total light in a non-parametric way and we find R$_e$ = 12.5 $\pm$ 2 arcsec (1.51 $\pm$ 0.24 kpc). 

In Fig.~3 we show the surface brightness profile of the galaxy in the g band. We have fit the profile with a Sersic function, again excluding the central star clusters. We adopt the Sersic fit values of n = 0.68, R$_e$ = 17.7 $\pm$ 0.5 arcsec (2.14 $\pm$ 0.06 kpc) and $\mu_e$ = 26.0 mag. per sq. arcsec. We note that if UDG1 is associated with NGC 5813 at a distance of 32 Mpc, then its size is closer to 
2.7 kpc. If located closer than 17.5 Mpc, the galaxy would no longer be regarded as a UDG (as defined by R$_e$ $>$ 1.5 kpc.) 
The central surface brightness of the galaxy is difficult to estimate due to the centrally located star cluster which dominates the profile. However extrapolating the outer profile to the inner regions we estimate $\mu_0$ =  24.8 mag. sq. arcsec in the g band.
The size and surface brightness at the effective radius (see Table 1) places the galaxy in the UDG category. 


After identifying UDG1 we found it was tabulated in the NGC 5846 group catalog of Mahdavi et al. (2005) as galaxy NGC~5846-156. They classified it as a ``very low surface brightness'' galaxy and measured a total magnitude of M$_R$ = --14.41 and R$_e$ = 10.5 arcsec (1.3 kpc at an assumed distance of 24.89 Mpc).
This identification occurred before the term ultra diffuse galaxy had been coined and falls below the nominal size limit of 1.5 kpc. 
It is not clear how they handled the compact sources around UDG1 when measuring the total magnitude or the size of the galaxy. We note that Mahdavi et al. (2005), in their 10 sq. deg imaging survey, identified another ten sources as very low surface brightness, but all of these sources have much smaller sizes than UDGs and so may be considered as low surface brightness dwarf galaxies (if they are indeed associated with the NGC 5846 group).

   \begin{table}
      \caption[]{NGC~5846$\_$UDG1 Properties}
         \label{KapSou}
            \begin{tabular}{lr}
            \hline
            \noalign{\smallskip}
            Property      &   Value\\
            \noalign{\smallskip}
            \hline
            \noalign{\smallskip}
            RA (J2000) & 15:05:20\\
            Dec (J2000) & 01:48:47\\
            m$_{g}$ (mag) & 18.0 $\pm$ 0.2\\ 
            R$_{e}$ (arcsec)& 17.7 $\pm$ 0.5\\
            g--i (mag) & $\sim$1\\
            $\mu_e$ (mag per sq. arsec) & 26.0  $\pm$ 0.05\\
            $\mu_0$ (mag per sq. arcsec) & 24.8 $\pm$ 0.1\\
            
            \noalign{\smallskip}
            \hline
\end{tabular}
   \end{table}

   \begin{figure}
   \centering
   \includegraphics[width=7cm,angle=0]{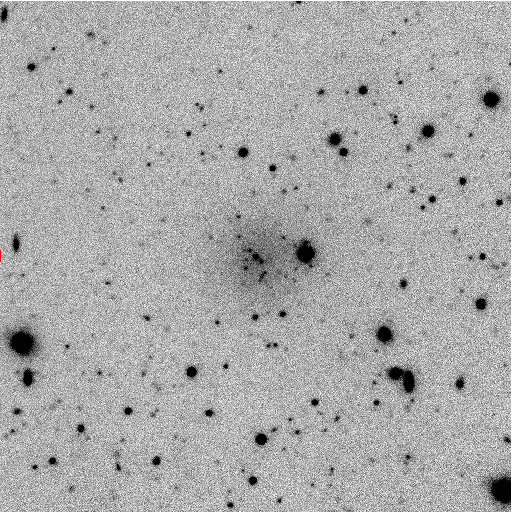}
      \caption{VEGAS image of UDG1. The image shows a 210$\times$210 arcsec g band image cutout of the larger VEGAS image, with North up and East left. UDG1, and its system of globular clusters, can be seen at the centre of the image. The centre of the NGC 5846 group is located to the South-West.     
              }
         \label{UDGimage}
   \end{figure}
   

We identify 20 compact objects projected near UDG1. We count up the objects within a 
region of 50 $\times$ 50 arcsec centred on the galaxy. This area corresponds to roughly twice the galaxy effective radius, which is typical of the GC systems of UDGs studied to date (see Forbes 2017). The surface density distribution of these sources within circular annuli about the galaxy centre is shown in Fig.~2. The distribution falls off with radius, which indicates that the sources (globular cluster candidates) are physically associated with UDG1. 

     \begin{figure}
   \centering
   \includegraphics[width=7cm,angle=0]{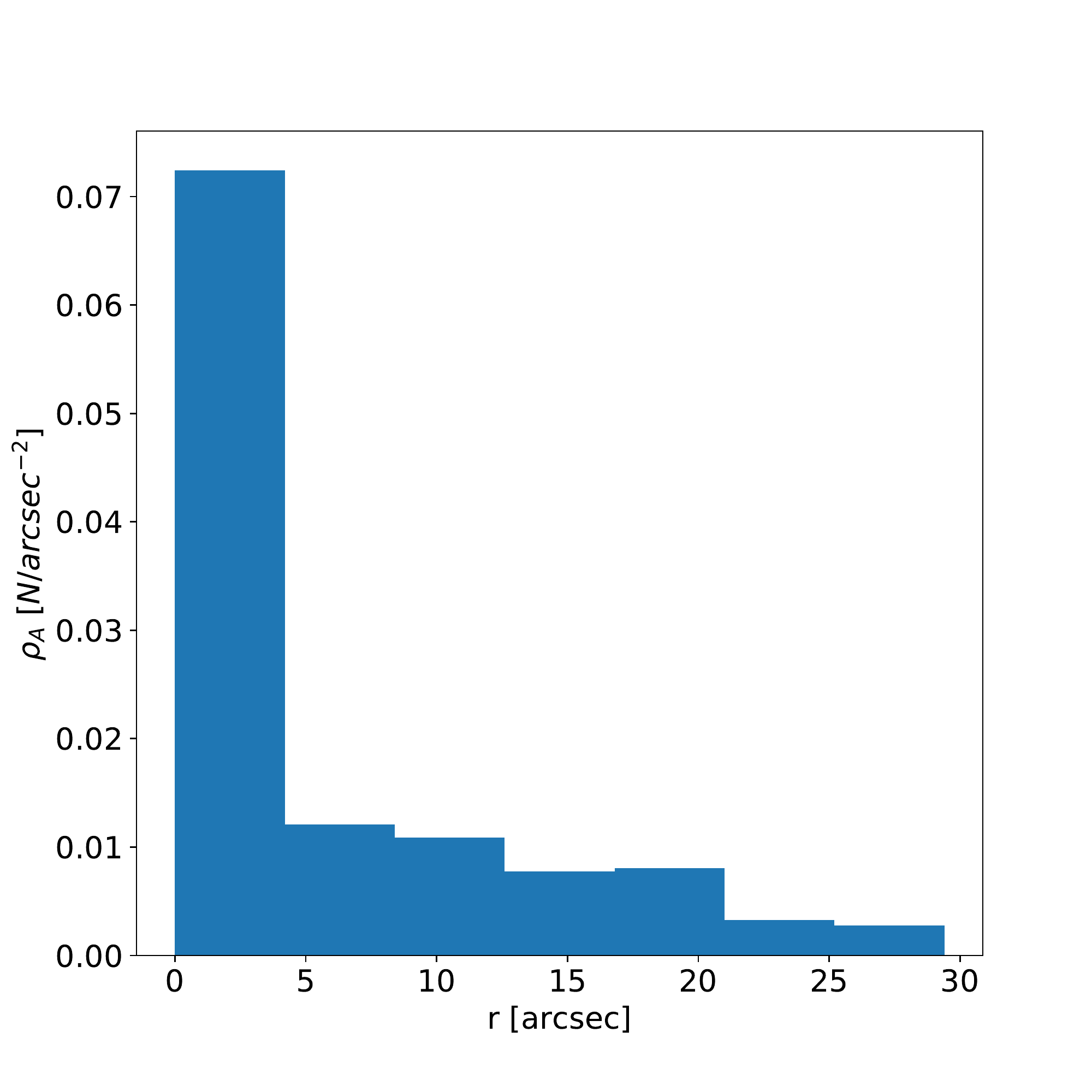}
      \caption{Surface density distribution of globular cluster candidates within a 50x50 arcsec box around UDG1. The surface density falls off with radius from the centre of the galaxy indicating that the sources are physically associated with UDG1. 
            }
         \label{SD}
   \end{figure}

Using {\it qphot} within IRAF, we 
measure the magnitude of all sources with sizes that are less than two times the seeing (in order to exclude large background galaxies).  Given that these compact sources are fairly crowded in the central regions, we measure the magnitude in a small aperture (radius of 4 pixels) with a local background subtraction  and apply an aperture correction based on measurements of a dozen isolated objects of similar brightness. The aperture correction to large radius is 0.63 $\pm$ 0.15. This error dominates our photometric uncertainty.  
The coordinates and total g band magnitudes for each compact source (i.e. GC candidates) within a 50 $\times$ 50 arcsec area are  listed in Table 2, including the central star cluster/nucleus GC9. 

Three of the objects in Table 2 have SDSS measured magnitudes and colours, i.e GC16, GC9 and GC19. They have SDSS g band magnitudes of 21.78, 22.25 and 23.36 respectively. GC16 has a colour of g-i = 1.24, which is very red for a GC and we suspect it is a foreground star or background galaxy. 
The other two have g-i $\sim$ 0.8 consistent with a GC.  
A few of the very brightest sources can just be seen in the shallow VEGAS i band image of the NGC 5846 group. From those sources we measure typical colours of g--i $\sim$ 0.9, which are consistent with those of GCs. 

In order to estimate the number of contaminants (e.g. background compact galaxies or foreground stars) which may be included in our list of UDG1 GC candidates, we identify several 50 $\times$ 50 arcsec regions in the outskirts of the full mosaic image. We estimate 3 $\pm$ 3 contaminant objects in our list of candidate GCs. Thus subtracting 3 from our 20 GC candidates, we estimate that 17 are bona fide GCs. 

In Fig.~4 we show a histogram of the g band magnitudes of the decected compact sources within 50 $\times$ 50 arcsec around UDG1. The globular cluster luminosity function (GCLF) in non-UDG dwarf galaxies can be well represented by a Gaussian log-normal distribution. For example, Miller \& Lotz (2007) found Virgo dEs to have M$_V$ = --7.3 $\pm$ 0.1 with a Gaussian width of $\sigma$ = 1.2 $\pm$ 0.1. Similar values were found by Georgiev et al. (2009) for local dwarf galaxies of various types. However, the GCLF parameters for UDGs are currently have a wide range of possible values.  
Peng \& Lim (2016) bravely fit a Gaussian to the luminosity distribution of their dozen GCs in the Coma UDG DF17, measuring M$_V$ = --7.37 $\pm$ 0.34 and $\sigma$ = 0.76 $\pm$ 0.23. van Dokkum et al. (2016) studied the richer GC system of DF44, also in Coma, and showed that Gaussians of 
M$_V$ = --7.5 and $\sigma$ = 1.2, and 
M$_V$ = --7.2 and $\sigma$ = 1.0 provided acceptable fits. In a later paper, van Dokkum et al. (2017) found $\sigma$ = 0.82 $\pm$ 0.16 to be the best fit to the combined GCLF of DF44 and another Coma UDG (DFX1). 
Given this large range of possible values for the width of the GCLF for UDGs, and the unknown distance to UDG1 (needed to set the absolute magnitude of the GCLF peak), the total number of GCs associated with UDG1 is poorly constrained. Here we prefer to quote the number of GCs detected, corrected for background contamination (i.e 17), as a lower limit to the total GC system. We note that for an assumed distance of 24.89 Mpc, a Gaussian GCLF with M$_V$ = --7.3 would peak at V = 24.7 or g $\sim$ 25.1 (for V = g -- 0.35; Capaccioli et al. 2015).




Again assuming V = g -- 0.35, UDG1 has an extinction-corrected  V band magnitude of M$_V$ = --14.5 and the lower limit on its globular cluster specific frequency is S$_N$ $>$ 27. 
This is high for a normal dwarf galaxy but not exceptional for a UDG (see 
Lim et al. 2018; Amorisco et al. 2018). 
We have included the central star cluster as a GC candidate in the above calculation, however UDG1 may be a nucleated UDG which are quite common in the Coma cluster (Lim et al. 2018). 


Burkert \& Forbes (2019) have recently extended the 
GC-halo mass scaling relation to lower masses, i.e. log~M$_h$ = 9.68 + 1.01~log~N$_{GC}$, where M$_h$ is the total halo mass in solar masses and N$_{GC}$ the number of GCs. Assuming the relation is valid for UDGs, our detection of at least 17 GCs implies log M$_h$ $>$ 10.92.  This halo mass lies between a dwarf galaxy and an L$^{\ast}$ galaxy like the Milky Way.
Future work on NGC~5846$\_$UDG1 should include attempts to obtain both the stellar velocity dispersion and, if possible, the radial velocities of individual GCs. Both of these measures will give a dynamical mass and hence indicate whether UDG1 resides in a massive dark matter halo as implied by its GC counts. 


   \begin{figure}
   \centering
   \includegraphics[width=7cm,angle=-90]{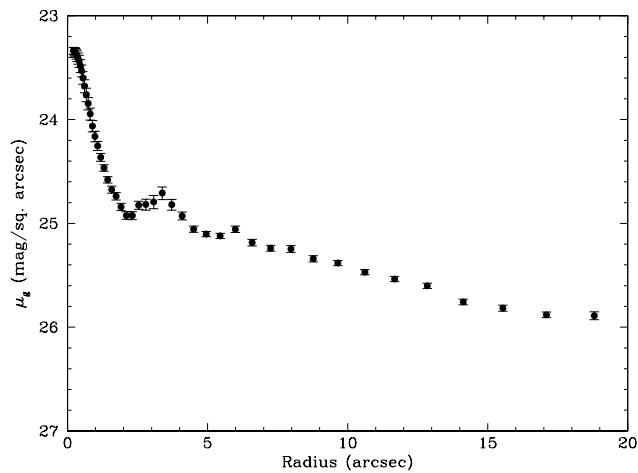}
      \caption{Surface brightness profile of UDG1 in the g band. Error bars show the formal measurement uncertainty. The inner regions are dominated by a bright compact source.
              }
         \label{sb}
   \end{figure}
%
   \begin{figure}
   \centering
   \includegraphics[width=7cm,angle=0]{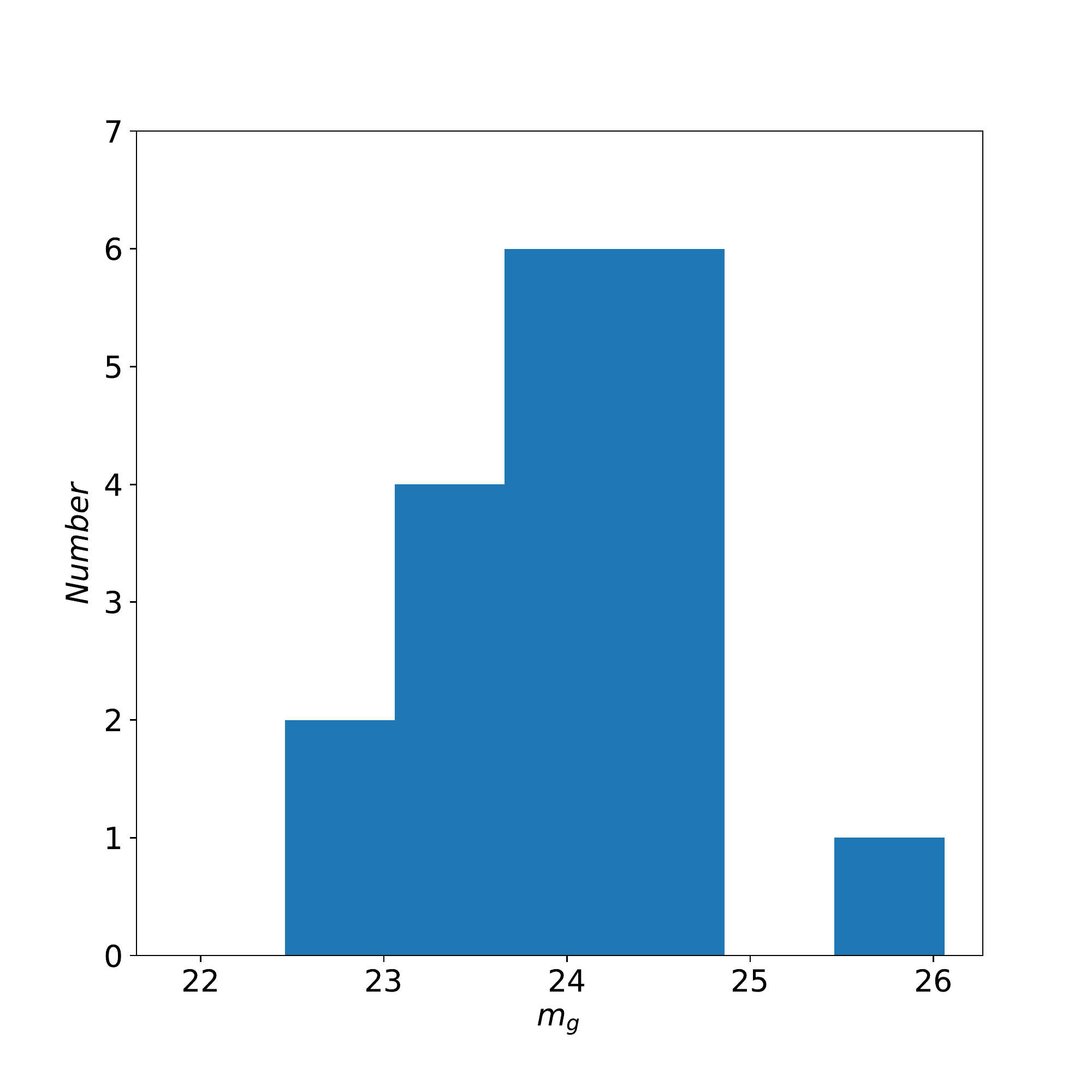}
      \caption{Histogram of g band magnitudes of compact objects associated with UDG1. 
      A Gaussian GCLF with M$_V$ = --7.3 would peak at V = 24.7 or g $\sim$ 25.1 for an assumed distance of 24.89 Mpc. 
              }
         \label{gclf}
   \end{figure}
   
   \begin{center}
\begin{table}[]
 \caption[]{UDG1 Globular Cluster Candidate Magnitudes}
\begin{tabular}{llll}
Object & RA         & Dec        & m$_{g}$ \\
 & (J2000) & (J2000) & (mag)\\
\hline
GC1 &15:05:21.5 & 1:48:54.8 & 23.96              \\
GC2 &15:05:21.0   & 1:49:02.0   & 24.85              \\
GC3 &15:05:20.8 & 1:49:03.0   & 23.28   \\
GC4 &15:05:18.9 & 1:49:06.6 & 24.31  \\
GC5 &15:05:19.5 & 1:48:54.4 & 23.72 \\
GC6 &15:05:20.6 & 1:48:41.8 & 23.38    \\
GC7 &15:05:20.5 & 1:48:45.3 & 24.44   \\
GC8 &15:05:20.4 & 1:48:49.2 & 23.58              \\
GC9 &15:05:20.3 & 1:48:46.7 & 22.49   \\
GC10 &15:05:20.1 & 1:48:44.7 & 22.89   \\
GC11 &15:05:19.6 & 1:48:39.9 & 24.81             \\
GC12 &15:05:19.5 & 1:48:38.1 & 24.46               \\
GC13 &15:05:19.2 & 1:48:41.5 & 24.23              \\
GC14 &15:05:19.9 & 1:48:34.1 & 24.79              \\
GC15 &15:05:18.9 & 1:48:39.8 & 25.51              \\
GC16 &15:05:19.6 & 1:48:23.0   & 21.77              \\
GC17 &15:05:18.8 & 1:48:42.7 & 23.69              \\
GC18 &15:05:20.0   & 1:48:39.9 & 23.64   \\
GC19 &15:05:20.1 & 1:48:38.4 & 23.69   \\
GC20 &15:05:20.2 & 1:48:36.9 & 24.07  \\
\hline
\end{tabular}
\end{table}
\end{center}

\section{Conclusions}

Here we present the discovery of an ultra diffuse galaxy which we assume to belong to the NGC 5846 group. We name this object NGC~5846$\_$UDG1. This galaxy was also present as the very low surface brightness galaxy NGC~5846--156 in the catalog of Mahdavi et al. (2005). They measured the magnitude and size of the galaxy but did not discuss the compact star clusters associated with it. 
At an assumed  distance of $\sim$25 Mpc for UDG1, we measure an effective radius of 2.1 kpc and a g band surface brightness at the effective radius of $\mu_e$ = 26.0 mag. sq. arcsec placing it in the UDG class. From the halo mass relation of Mancera Pina et al. (2018) up to half a dozen more UDGs may be expected to be found associated with the NGC 5846 group.  
We find a number of compact sources that have magnitudes and a spatial distribution consistent with being globular clusters associated with UDG1. After statistically correcting for background objects, we place a lower limit on the system to be 17 GCs (one of which may be a central star cluster/nucleus). 
Our lower limit on the GC system suggests a halo mass in excess of 8$\times$10$^{10}$ M$_{\odot}$.  This is in stark contrast to the galaxy's stellar mass of $\sim$10$^8$ M$_{\odot}$.  Spectroscopic followup is required to confirm its membership of the NGC 5846 group, measure its internal kinematics and stellar population properties. We expect many more ultra diffuse galaxies to be discovered in the deep and wide VEGAS survey images of massive nearby early-type galaxies.

\begin{acknowledgements}
      We thank the full VEGAS survey team. DF thanks the ARC via DP160101608. We thank the referee for a rapid and useful report.
      MS acknowledges financial 
      support from the VST project (P.I. P. Schipani).
This work is based on visitor mode observations taken at the  ESO  La  Silla  Paranal  Observatory  within  the  VST  Guaranteed  Time  Observations,  Programme  ID 095.B-0779(A). NRN acknowledges financial support from the “One hundred talent program of Sun Yat-sen University” and from the European Union Horizon 2020 research and innovation programme under the Marie Skodowska-Curie grant agreement n. 721463 to the SUNDIAL ITN network.

\end{acknowledgements}


\end{document}